# Disaster Management in the Era of Agentic AI Systems: A Vision for Collective Human-Machine Intelligence for Augmented Resilience

Bo Li[1], Junwei Ma[1*], Kai Yin[1], Yiming Xiao[1], Chia-Wei Hsu[1], Ali Mostafavi[1]

1 Urban Resilience.AI Lab, Zachry Department of Civil and Environmental Engineering, Texas A&M University, College Station, Texas, United States.

* Corresponding author: Junwei Ma, E-mail: jwmatamu@gmail.com.

## ABSTRACT

The escalating frequency and severity of disasters routinely overwhelm traditional response capabilities, exposing critical vulnerability in disaster management. Current practices are hindered by fragmented data streams, siloed technologies, resource constraints, and the erosion of institutional memory, which collectively impede timely and effective decision making. This study introduces "Disaster Copilot", a vision for a multi-agent artificial intelligence system designed to overcome these systemic challenges by unifying specialized AI tools within a collaborative framework. The proposed architecture utilizes a central orchestrator to coordinate diverse "sub-agents", each specializing in critical domains such as predictive risk analytics, situational awareness, and impact assessment. By integrating multi-modal data, the system delivers a holistic, real-time operational picture and serve as the essential AI backbone required to advance Disaster Digital Twins from passive models to active, intelligent environments. Furthermore, it ensures functionality in resource-limited environments through on-device orchestration and incorporates mechanisms to capture institutional knowledge, mitigating the impact of staff turnover. We detail the system architecture and propose a three-phased roadmap emphasizing the parallel growth of technology, organizational capacity, and human-AI teaming. Disaster Copilot offers a transformative vision, fostering collective human-machine intelligence to build more adaptive, data-driven and resilient communities.

# 1. Introduction

Disasters, such as hurricanes, floods, or wildfires, are striking with increasing frequency and severity worldwide (Raymond et al., 2020). Future projections suggest that extreme weather events will intensify further, creating pressing need for more effective disaster management (Camps-Valls et al., 2025). In parallel, artificial intelligence (AI) has advanced rapidly, with growing capabilities in prediction, sensing, and decision support. These advances align directly with the escalating demands of disaster management, positioning AI to transform how we anticipate, respond to, and recover from disasters.

Over the past decade, artificial intelligence has demonstrated remarkable potential in advancing disaster management. AI enables the fusion of diverse and large-scale data sources, such as satellite imagery, social media, sensor networks and historical records, into rapid analytics that enhance situational awareness. Building on this foundation, machine learning techniques provide more accurate hazard forecasting, enable emerging risk detection, and accelerate impact assessment by identifying patterns from extensive data. Furthermore, AI supports creation of digital twins, virtual replicas of physical regions that can simulate disaster scenarios and test mitigation strategies. These digital environments allow practitioners to evaluate trade-offs, design more resilient infrastructure, and embed lessons from past events into evolving models, thereby strengthening long-term disaster resilience. Collectively, these capabilities establish AI as a critical partner to support all phases of disaster management.

Despite significant progress, the integration of AI into disaster management remains constrained by several systemic challenges. First, while specialized AI models excel at specific tasks, their growing proliferation has not adequately resolved information and workflow fragmentation, while in many cases, has amplified it. Current AI models are typically designed for highly specialized functions, such as forecasting hazards, detecting infrastructure damage, or predicting evacuation demands. Within these domains, they often achieve high levels of accuracy and timeliness. However, their outputs are generated in heterogeneous formats, with different spatial and temporal resolutions, update frequencies, and underlying assumptions. These inconsistencies make it difficult to align results across agencies such as meteorological services, utilities, public health departments, and emergency logistics teams. Rather than converging into a shared analytical framework, the outputs of these models often form new data silos and coordination gaps. For example, a flood forecast that identifies inundates areas may not be readily aligned with transportation network models, making it challenging to determine which evacuation routs remain safe and accessible.

The fragmentation problem is further compounded at the organizational level. Disaster management requires the seamless collaboration of diverse actors. Without a federated platform, communications among these teams could be dispersed across emails, phone calls, and stand-alone dashboards, slowing the flow of information, and disrupting coordination. For example, headquarters may allocate shelters based on regional forecasts, while on-site responders facing unexpected local flooding redirect evacuees to different sites, leading to unbalanced shelter allocation and less prompt disaster response.

This lack of interoperability, stemmed from the fragmentation of AI models, creates not only technical obstacles but also significant cognitive burdens. During unfolding disaster events, emergency managers must navigate multiple dashboards, interpret highly specialized outputs, and integrate disparate insights into actionable strategies under tight time pressure and uncertainty. Such integration requires considerable expertise and effort, which can undermine the speed and precision of decision making. As disasters grow more frequent and complex, these fragmented workflows risk causing information overload and limiting the full value of AI-generated insights, which creates operational bottleneck for effective disaster management. Addressing this challenge calls for a federated mechanism that connects specialized models with a coordinated and interoperable framework, to ensure that their collective strengths translate into timely, consistent, and holistic support for effective crisis response.

Furthermore, current AI tools inadequately address the problem of preserving institutional knowledge and operational experience. Emergency management agencies often operate with limited resources and bandwidth (Farhana & Siti-Nabiha, 2024; Opoku-Boateng, Agyei, Asibey, & Mintah, 2024), while the knowledge database they must rely on is vast, scattered and frequently updated. This mismatch makes it difficult for practitioners to systematically capture, organize and apply lessons learned from historical disaster events. In practice, agencies often depend heavily on the personal experience of staff, which becomes a vulnerability when personnel turnover occurs. As experienced professionals depart, they take with them critical insights into local hazards, community vulnerabilities, and prior response strategies, leaving successors without access to accumulated knowledge (L. Li et al., 2024; Mathews, Vickery, & Peek, 2024).. Without mechanisms to retain and transfer expertise, agencies struggle to apply lessons consistently, leading to a cycle of reactive rather than adaptive response. As a result, best practices and localized risk knowledge remain underutilized, reducing the effectiveness of disaster management efforts. Addressing this gap requires a platform that can capture institutional memory, consolidate dispersed guidance, and make knowledge both retrievable and actionable for future events.

The point of departure for this study is the critical recognition that, despite substantial progress in specialized applications, AI tools in disaster management remain fragmented and largely isolated. This persistent lack of interoperability creates technical inefficiencies and places heavy cognitive demands on decision-makers, who must reconcile disparate insights under conditions of urgency and uncertainty. Compounding these issues is the erosion of institutional knowledge driven by resource constraints and frequent staff turnover, which prevents lessons learned from being systematically captured and applied across events. Taken together, these challenges reveal that the critical barrier in disaster management is not the absence of advanced AI capabilities but the absence of a unifying framework to integrate and sustain them.

In light of these challenges, this study introduces a significant paradigm shift by proposing the "Disaster Copilot", a novel multi-agent AI architecture designed to orchestrate diverse specialized AI tools into a cohesive ecosystem, thereby enabling the collective human-machine intelligence necessary for truly adaptive and effective disaster response (Fig.1). In this architecture, each "sub-agent" serves as a specialist for a given domain, while a top-level orchestration component coordinates their actions and exchanges their outputs. Rather than forcing every AI model or process into a single, monolithic platform, this multi-agent AI approach preserves each system's specialized strengths, yet ensures they interoperate seamlessly to provide a cohesive, end-to-end view of disaster scenarios. By designing a collaborative network of agents that communicate through well-defined interfaces, emergency managers and key decision-makers can efficiently access all relevant insights via one orchestrated system. Moreover, the multi-agent framework is explicitly designed to support human-AI teaming, ensuring that as personnel change over time, the knowledge embedded within the agents is preserved, thereby providing continuity and sustain institutional memory. The orchestration layer also reduces information overload by filtering and prioritizing updates, letting responders focus on high-impact decisions. Taken together, Disaster

Copilot has provided a transformative blueprint for significantly enhancing disaster resilience and responsiveness in an era of escalating global threats.

This paper will unfold this vision by first reviewing the current state of AI in disaster management to highlight existing capabilities and gaps (Section 2). We will then detail the proposed multi-agent AI system architecture, illustrating its components and interactions (Section 3). Subsequently, we will present a realistic scenario to demonstrate the system in action (Section 4), followed by a three-phased roadmap for its research, development, and institutional integration (Section 5). We will discuss the broader implications of this approach for transforming disaster management into a more adaptive, collaborative, and resilient endeavor (Section 6), and conclude with key insights (Section 7).

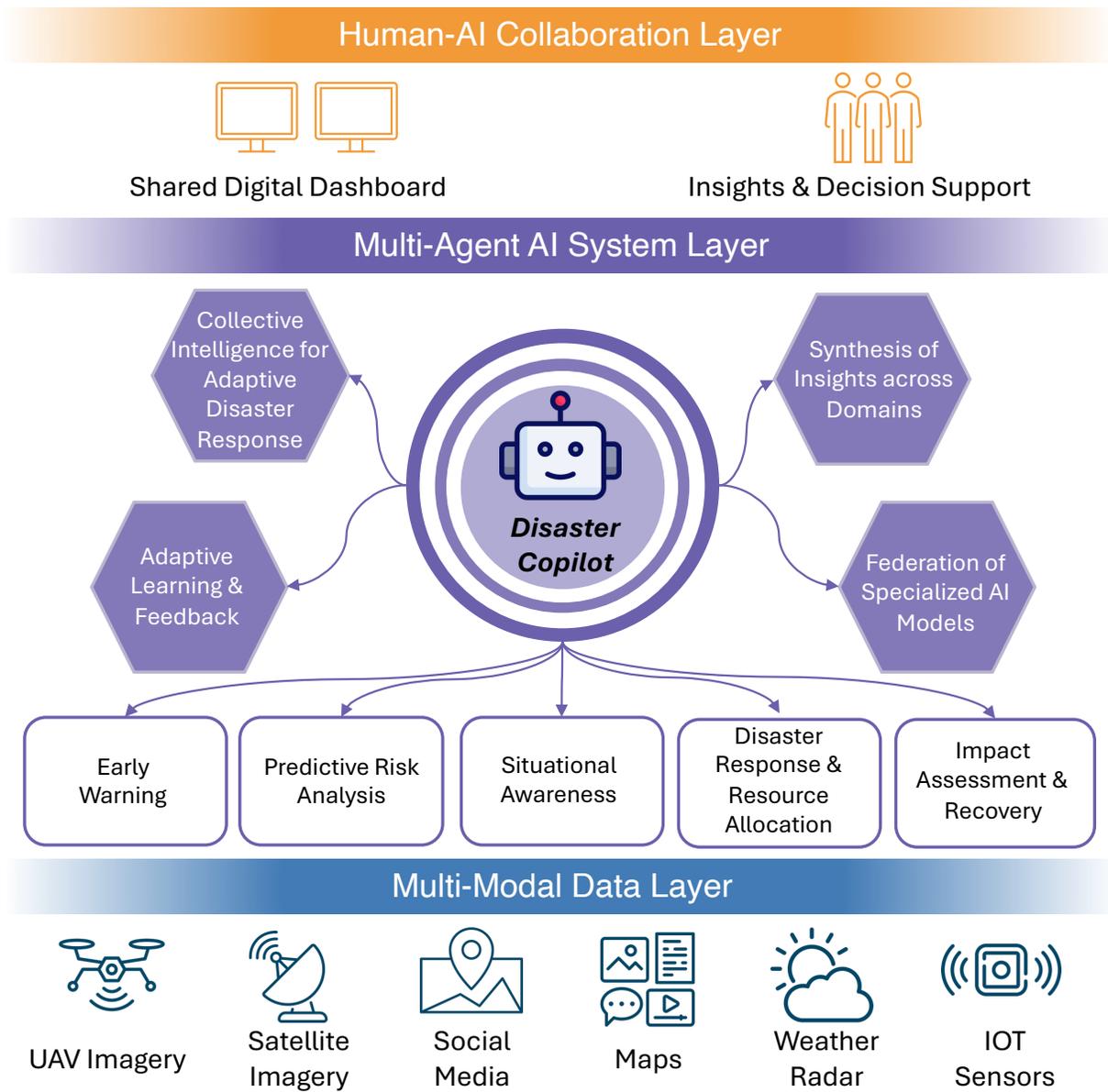

**Fig 1. Conceptual framework of Disaster Copilot: a layered architecture for collective human-machine intelligence.** This diagram illustrates the integration pipeline across multiple layers. It begins with the ingestion of diverse multi-modal data (e.g., satellite imagery, spatial datasets, crowdsourced data), which is processed by Specific AI Models focused on tasks such as hazard mapping, risk assessment, and situational awareness. The multi-agent AI Systems layer, featuring the Disaster Copilot, orchestrates these specialized tools and synthesizes their outputs. The process culminates at the Human-AI Collaboration layer, where the synthesized intelligence

provides a unified operational picture to support coordinated decision-making and augmented resilience.

## 2. AI for Disaster Management: State of Knowledge and Practice

In this section, we present a range of AI applications that have already seen development or deployment in disaster contexts. Each subsection highlights key methods, data sources, and real-world examples demonstrating how AI can enhance different aspects of emergency and risk management.

### 2.1 Early warning

AI techniques, particularly machine learning and deep learning, are transforming natural disaster early warning systems by enhancing the speed, accuracy, and efficiency of predictions, especially where physics-based models are computationally intensive (Liao, Wang, Chen, & Lai, 2023; J. Liu, Lee, & Zhou, 2023). Leveraging diverse datasets such as satellite imagery, meteorological observations, seismic activity, and wireless sensor networks, AI models can uncover complex patterns and anomalies that precede hazards (Charan et al., 2024; Y. Zhang, Geng, Sivaparthipan, & Muthu, 2021).Recent research highlights several important directions. In flood forecasting, deep learning models such as CNNs and LSTMs are being developed to simulate inundation dynamics and predict streamflow with high accuracy, even under data-scarce conditions (Liao et al., 2023; Ma et al., 2024). In wildfire prediction and detection, hybrid approaches combining LSTM, CNN, RF, and SVM leverage environmental and remote sensing data to anticipate ignition risk and classify burn areas (Bakhshian & Martinez-Pastor, 2023; Lakshmanaswamy, Sundaram, & Sudanthiran, 2024). AI has also advanced coastal hazard forecasting, with CNN-based methods integrating wind field and water level data to improve storm surge predictions (Xie, Xu, Zhang, & Dong, 2023). Similarly, in seismic early warning, optimized ensemble models such as XGB use real-time seismic and buoy data to estimate intensity within seconds, enabling faster tsunami and earthquake alerts (Abdalzaher, Soliman, & El-Hady, 2023). Collectively, these studies demonstrate a clear trend toward applying AI across diverse hazard domains, with a focus on fusing multi-source data and designing task-specific architectures that extend lead times and improve predictive reliability.

## 2.2 Predictive risk analysis

AI-driven predictive models are increasingly used to anticipate where, how, and when hazards may occur, supporting the creation of risk maps that identify vulnerable communities and infrastructure. Machine learning and deep learning methods enhance prediction capabilities, as CNNs, LSTMs, and hybrid or ensemble models have been applied to floods, wildfires, and tropical cyclones, achieving high accuracy in susceptibility mapping, streamflow forecasting, and flood damage prediction(Bui et al., 2023; Maha Arachchige & Pradhan, 2025; Moayedi & Khasmakhi, 2023; Yongyang Wang, Zhang, Xie, Chen, & Li, 2025; J. Yang et al., 2024). In this process, AI is increasingly used to automate workflows by integrating heterogeneous geospatial, environmental, and socioeconomic datasets, and accelerating computationally intensive models (e.g. large-scale surge simulations) to make risk mapping more efficient and operationally usable (Eini, Kaboli, Rashidian, & Hedayat, 2020; Habibnia & van de Lindt, 2025; Y. Liu et al., 2021; H. Singh, Ang, & Srivastava, 2025). Beyond mapping, researchers have leveraged explainable AI to increase transparency and trust in model outputs, with methods such as SHAP and Partial Dependence Plots being applied to reveal key vulnerability drivers across contexts (Darabi et al., 2019; C. Liu & Mostafavi, 2025; Maha Arachchige & Pradhan, 2025). Also, generative approaches such as GANs and diffusion models are explored to augment scarce or imbalanced datasets, impute missing observations, and better represent rare or extreme events (Al-Najjar & Pradhan, 2021; Do Lago et al., 2023; Durall, Ghanim, Fernandez, Ettrich, & Keuper, 2023).

## 2.3 Situational awareness

Disaster situation awareness refers to the ability to gather, process and interpret diverse, real-time information on hazards, infrastructure, and population in order to support timely and effective decision-making under uncertainly (Comfort, Ko, & Zagorecki, 2004; Vieweg, Hughes, Starbird, & Palen, 2010). AI is advancing this capability through multiple pathways. Multimodal fusion and relevance filtering have been developed to handle the noise and heterogeneity of social media, combining geospatial models, image classifiers, NLP-based text analysis, and user reliability measures to extract high-value signals during crises such as Hurricane Irma (Mohanty et al., 2021) and Hurricane Laura (Zhou, Kan, Huang, & Silbernagel, 2023). Crowdsourced social sensing extends official monitoring by geocoding flood-relevant tweets in Mumbai to detect overlooked hotspots and mobility impacts (Navalkar et al., 2025), fusing Twitter with Waze alerts to capture community needs (Salley, Mohammadi, Xie, Tien, & Taylor, 2024), and applying multimodal deep learning to integrate text, images, and geospatial data for comprehensive public safety monitoring (Sangeetha et al., 2024). AI-driven remote sensing and sensor networks enhance disaster observability, applying Bayesian network modelling for optimal flood gauge placement (Farahmand, Liu, Dong, Mostafavi, & Gao, 2022), deep learning to fuse SAR imagery with social media for flood mapping (Sadiq, Akhtar, Imran, & Ofli, 2022) and automated change detection on SAR backscatter for rapid landslide assessment (Handwerger et al., 2022). Edge and aerial computing contribute by enabling UAVs equipped with compressed CNNs to classify disaster scenes such as collapsed buildings or fires in real time (Ijaz et al., 2023). The recently thriving large language models add new capabilities, with instruction-tuned variants like CrisisSense-LLM improving the classification of disaster-related tweets (Yin, Li, Liu, Mostafavi, & Hu, 2024) and human-centered systems like CROMEx supporting incident detection and practitioner-guided model selection (Senarath, Mukhopadhyay, Purohit, & Dubey, 2024). Finally, integrity in crisis communication is being reinforced through misinformation-aware community detection frameworks, employing deep learning–based mis-information detection to filter unreliable content during emergencies (Apostol, Truică, & Paschke, 2024).

## 2.4 Disaster response and resource allocation

Effective disaster response relies on anticipating demand, coordinating scarce resources, and identifying needs in a timely manner, and AI is increasingly being applied to support these functions. Recent studies illustrate how AI enhances traffic management during evacuations, with LSTM-based models predicting traffic demand up to 24 hours in advance by combining sensor and social media data (Roy, Hasan, Culotta, & Eluru, 2021),while GIS-driven simulations refine evacuation time estimates in wildfire-prone resort areas (D. Li, 2022). In parallel, UAV-based deep learning systems such as YOLOv5 enable rapid victim detection in complex terrains, significantly reducing search and rescue times (Caputo et al., 2021; Xing et al., 2022). AI is also improving resource allocation and logistics. Machine learning frameworks accurately forecast relief demand following earthquakes (Biswas, Kumar, Hajiaghaei-Keshteli, & Bera, 2024), deep reinforcement learning optimizes real-time distribution during storm surge emergencies (Yuewei Wang, Chen, & Wang, 2023), and transfer learning models anticipate cross-disaster commodity needs under uncertainty (Zheng et al., 2021). By blending historical data with live hazard predictions, AI systems help emergency managers proactively position resources to ensure more rapid and effective relief to those most in need.

**2.5 Impact assessment and recovery**

Following major disasters, decision-makers face urgent demands for reliable intelligence on damages, disruptions and recovery prospects, and AI has been offering tools across multiple domains. For example, deep learning models have been applied to satellite and UAV images to automate large-scale mapping and multi-class classification of building damages (Berezina & Liu, 2022; Braik & Koliou, 2024; L. Zhang & Pan, 2022). Beyond assessing structural damages, AI techniques as been applied to automate impact assessment on mobility systems, with Bayesian CNNs estimating storm-related debris volumes (Cheng, Behzadan, & Noshadravan, 2024), and advanced unsupervised deep learning frameworks detecting fallen trees from satellite images, reducing reliance on manual surveys and enabling near real-time mobility assessments (Gazzea et al., 2021). Multimodal AI systems can fuse social media and hazard data to provide fine-grained assessment of community-level disaster impact in real time (Hao & Wang, 2021). Moreover, AI has been integrated to diagnose lifeline facility disruptions, anatomize recovery patterns and guide restoration patterns. For example, graph neural networks have been applied to reveal how inter-firm dependencies shape post-flood business recovery (Yang, Ogawa, Ikeuchi, Shibasaki, & Okuma, 2024), time series clustering of historical power outage records has captured fundamental resilience archetypes in utility restoration (B. Li & Mostafavi, 2024), and lightweight UAV-based CNNs have been deployed to guide rapid inspection and restoration of utility poles (Jeddi, Shafieezadeh, & Nateghi, 2023). These advances demonstrate how AI can deliver faster, finer-grained and more comprehensive impact insights to support adaptive recovery planning.

Collectively, these AI applications offer tangible benefits across all stages in disaster management. They demonstrate how machine learning, computer vision, natural language processing, and optimization solutions can enhance workflows that were previously manual. By accelerating and automating tasks such as hazard early detection, disaster situation awareness, and relief resource allocation, AI enables faster, more data-driven, and more adaptive decision-making under conditions of uncertainty.

Despite the rapid expansion of AI tools tailored to different disaster management tasks, these systems frequently operate as standalone applications, each with its own data sources, models, and user interfaces. While such specialization yields highly tailored solutions, it also reinforces fragmentation, requiring emergency managers to navigate multiple platforms and heterogeneous data formats under significant temporal pressure. In large-scale crises where effective decisions hinge on holistic, real-time information exchange, this lack of interoperability can cause critical lags and gaps. For example, a flood forecasting system may generate high-resolution inundation maps that remain siloed from evacuation route planning models, while social media-driven situation awareness outputs often fail to interface with resource allocation framework in a timely manner. These limitations underscore the pressing need for an orchestration mechanism capable of integrating heterogeneous AI applications into a coherent, interoperable ecosystem, ensuring that insights generated across domains can be synthesized and acted upon in real time.

## 2.6 Digital twin framework in disaster management

The emergence of Digital Twin technologies has been heralded as a transformative development in disaster management, offering the potential to create high-fidelity virtual replicas of urban environments and critical infrastructure (Fan, Zhang, Yahja, & Mostafavi, 2021). By integrating diverse data sources such as GIS, 3D modeling and real-time IoT sensor feeds, Digital Twins enable unprecedented capabilities for visualizing risks and running complex simulations of hazards such as hurricanes or wildfires (Braik & Koliou, 2023; Lewis et al., 2024). However, a critical review of existing frameworks reveals that many contemporary Digital Twins function primarily as sophisticated data aggregation and visualization platforms rather than truly intelligent systems. Their capacity for autonomous reasoning and dynamic analysis is often limited, relying heavily on pre-configured scenarios and intensive manual interpretation by experts (El-Agamy, Sayed, AL Akhatatneh, Aljohani, & Elhosseini, 2024). This reliance on human intervention to synthesize data and derive actionable insights fundamentally constrains the utility of Digital Twins during rapidly evolving crises, where the speed and complexity of events quickly outpace manual analytical capacities.

Furthermore, the analytical models embedded within current Disaster Digital Twins often suffer from the same operational fragmentation that characterizes standalone AI tools. While a Digital Twin might incorporate advanced simulations, these models frequently operate in isolation. For instance, a hydrological simulation predicting flood extent may not dynamically interface with a traffic model to update evacuation routes or with a logistics optimizer to reallocate resources in real-time. This lack of interoperability means that the Digital Twin cannot autonomously reason about cascading impacts or generate holistic response strategies. The critical gap, therefore, is not the fidelity of the virtual representation, but the absence of a unifying cognitive architecture, an orchestration mechanism capable of integrating these disparate models and data streams. Without this intelligent backbone, Digital Twins remain largely passive environments, unable to deliver the adaptive, integrated decision support required for managing modern disasters.

To address the limitations, this study proposed "Disaster Copilot", a multi-agent AI architecture to orchestrate diverse specialized AI tools into an integrated and interoperable ecosystem. In the next sections, we will elaborate the key architectural components of the proposed Disaster Copilot system, outlining the functions of individual agents and the orchestration strategies that enable their integration. This architecture provides a pathway for emergency management stakeholders to incorporate state-of-the-art AI solutions while preserving the specialized capabilities that underpin their value. Rather than introducing another isolated technological layer, the system establishes a modular yet unified ecosystem in which each specialized AI contributes to a comprehensive, adaptive, and resilient disaster management process.

## 3. Proposed multi-agent AI system architecture

Disaster management hinges on the ability to rapidly assess unfolding situations, synthesize diverse data, and make decisions that minimize impacts. Current AI solutions excel at specific tasks, such as damage assessment, risk forecasting, or resource allocation, but lack a unifying framework that allows them to collaborate seamlessly. By leveraging a Multi-modality Large Language Model (MLLM) as a central orchestrator, this multi-agent AI system integrates multiple specialized agents under one cohesive "disaster-copilot". The result is an ecosystem where domain experts (sub-agents specialized in particular tasks) can communicate and coordinate, thus enhancing each other's outputs and streamlining human decision-making. The proposed multi-agent AI system (Fig 2) incorporate the multiple key features: (1) federation of AI tools. In the system, each sub-agent addresses a specific aspect of disaster management (e.g. hazard mapping, risk analysis, external knowledge retrieval), while these sub-agents interact via well-defined interfaces managed by the system's task planner. The task planner acts as a central orchestrator, which decomposes complex objectives into discrete, domain-specific tasks, and then assigns them to the appropriate sub-agents, and integrate their outputs into a coherent operational picture. (2) full-modality, multilingual support. The system incorporates full-modality capabilities, enabling it to ingest, interpret and fuse diverse data types into a unified analytical workflow. This is particularly valuable in disaster management, where practitioners must process structured reports, sensor feeds, social media images, aerial videos, and verbal situation updates in near real time. In addition, the system offers multilingual processing, supporting both automatic translation and direct native language understanding across a broad set of languages. These capabilities ensure that critical information can be accessed and acted upon without language barriers among diverse stakeholders in disaster response. (3) The system offers scalability and deployment flexibility to meet the demanding, resource-constrained conditions of disaster management. Through model quantization and on-device orchestration, it can run efficiently on low-power hardware, enabling deployment at the disaster edge where connectivity and infrastructure may be compromised. This adaptability ensures that critical analytics remain available for both centralized command centers and field teams, supporting continuous decision-making throughout all phases of response. (4) The system incorporates a distributed, vectorized memory bank that preserves continuity of context across interactions and over extended crises. This persistent memory allows the system to track ongoing tasks, recall prior analyses, and maintain situational awareness even as conditions evolve or personnel rotate during prolonged crises. By retaining institutional knowledge, past decisions, and multi-step task progress, the system helps mitigate challenges associated with staff turnover and ensures that critical insights are not lost between personnel shifts.

The following subsections detail each major component of the multi-agent system, describing its purpose, technical underpinnings, and role in providing a robust disaster-copilot experience.

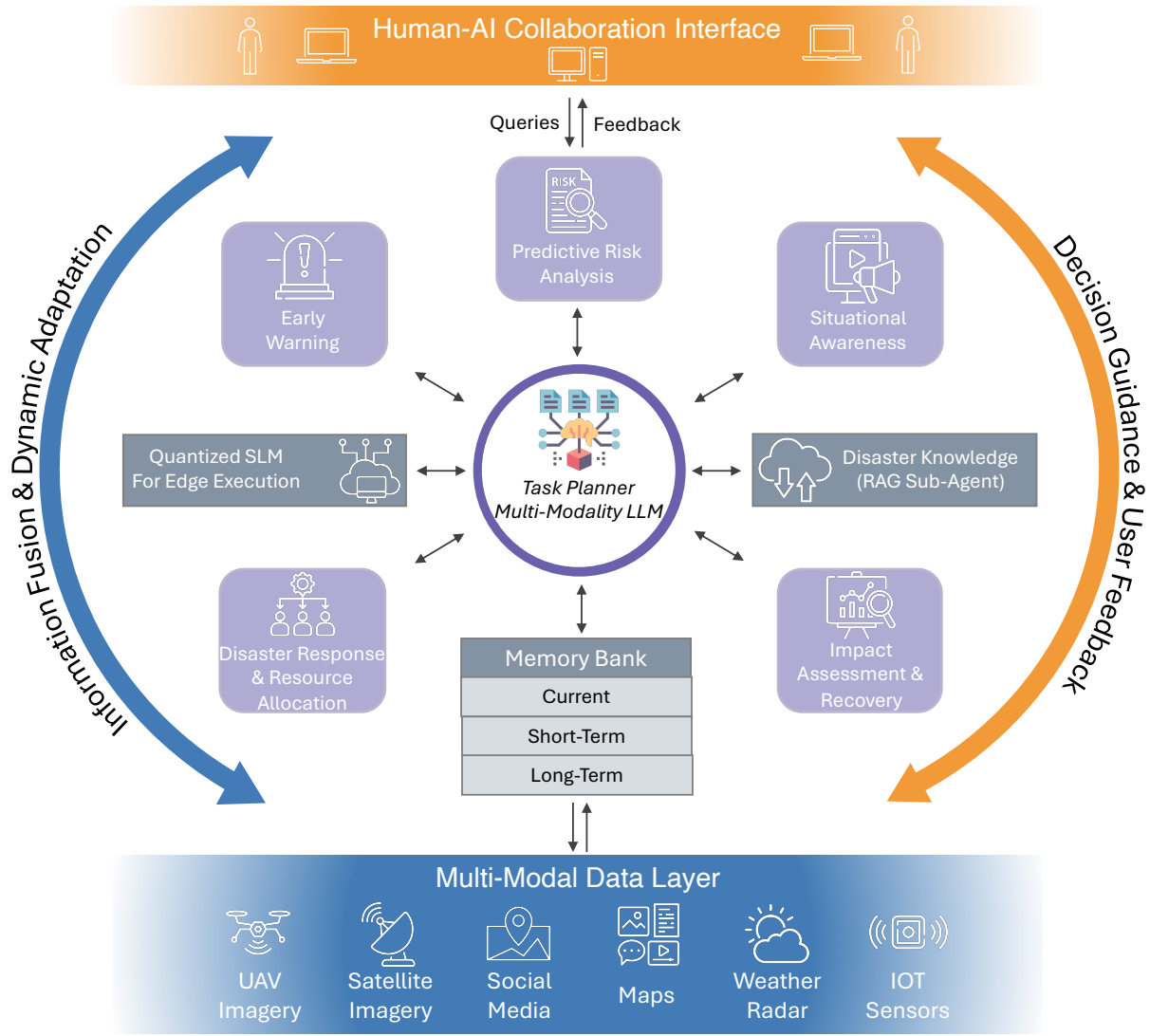

**Fig 2. Overall architecture of the propose multi-agent AI system.** The Task Planner (Multi-Modality LLM) serves as the central coordination hub of the Disaster Copilot, connecting and orchestrating all specialized sub-agents to enable adaptive and intelligent disaster management. Around the Task Planner, a network of specialized sub-agents performs complementary functions, including predictive risk analysis, early warning, situational awareness, impact assessment & recovery, and disaster response & resource allocation, collectively supporting data-driven situational intelligence and operational decision-making. The Task Planner is tightly integrated with two critical system components: the disaster knowledge (RAG) sub-agent, which retrieves and anchors AI reasoning to authoritative, up-to-date disaster knowledge bases; and the memory bank, which maintains current, short-term, and long-term contextual memory for sustained reasoning across events. Multi-Modal Data supplies heterogeneous inputs from UAV imagery, satellite observations, maps, social media, weather radar, and IoT sensors. The system is further enhanced by a quantized SLM for edge execution, enabling lightweight, on-device reasoning for field operations. Human-AI collaboration interface connects analysts and responders through bidirectional query and feedback loops, allowing real-time interaction with the AI system. Together, these components form a continuous cycle of information fusion, reasoning, and decision guidance, enabling Disaster Copilot to deliver evidence-grounded, context-aware, and dynamically adaptive support throughout disaster response and recovery.

## 3.1 Domain-specific task planner

A core element of the system is the central domain-specific task planner, which orchestrates complex user queries by decomposing them into actionable sub-tasks. Disaster management inherently involves multiple interdependent processes, such as hazard mapping, risk estimation, logistics planning, and situational communication, which cannot be handled effectively by isolated models. A coordinating mechanism is required to schedule tasks, manage dependencies, and combine outputs from specialized AI and machine learning models into a coherent response. The system addresses this need by employing a multi-modality large language model (MLLM) as the planner, leveraging its advanced capabilities in understanding, reasoning, generation, and interaction (Shen et al., 2023). Technically, the planner ingests full-modality user inputs, such as text, images, audio, and video, and interprets them as mission objectives that can be translated into structured workflows. Each sub-agent registers a structured description of its function, inputs, and outputs, which the MLLM incorporates into its context to generate a dependency graph of subtasks. This planning process is further refined through reinforcement learning from human feedback (RLHF), which improves the MLLM's ability to parse objectives, reason about the available sub-agents, and generate task plans that align with expert decision-making standards (Chaudhari et al., 2024). Execution results are validated against predefined schemas, stored in the memory bank, and summarized into human-readable outputs, ensuring transparency, adaptability, and reliability in dynamic disaster environments.

## 3.2 Domain professional sub-agents

Within the multi-agent AI system, the core tasks of disaster management are handled by domain professional sub-agents, each responsible for a cluster of AI models specialized in areas such as risk analytics, early warning, damage assessment, and resource allocation. These sub-agents serve as the operational bridge between high-level user queries and specialized analytical processes, translating general instructions into targeted computations and actionable outputs. Each sub-agent exposes its capabilities through standardized interface, specifying its function, required inputs, output schemas, modality support, and computational constraints, so that the central planner can select and invoke the most relevant sub-agents.

### 3.2.1 Multi-model approach within each sub-agent

Each domain-specific sub-agent employs a multi-model approach, integrating complementary machine learning and statistical techniques to address the complexity and uncertainty inherent in disaster management. Instead of relying on a single predictive or analytical model, sub-agents combine multiple models that specialize in different aspects of a task or operate at varying spatial and temporal resolutions. Sub-agents can dynamically select or weight models based on contextual factors such as hazard type, available data sources, or computational constraints at the edge. For example, a sub-agent aiming to perform flood prediction might maintain a convolutional LSTM for rainfall-runoff forecasting and a separate Bayesian network for coastal storm surge scenarios.

We design a prototype of each sub-agent specializing in different phases of disaster management, ensuring that the system covers the full cycle from early warning to post-disaster recovery. The Early Warning Sub-Agent employs time-series forecasting models such as LSTM and GRU, along with anomaly detection on real-time sensor signals and computer vision applied to satellite and camera imagery, to provide timely alerts for floods, wildfires, and other hazards. The Damage Assessment Sub-Agent integrates deep learning models for image segmentation (e.g., U-Net, Mask R-CNN) and change detection techniques applied to satellite, aerial, and drone imagery, enabling rapid identification of collapsed buildings, infrastructure losses, and spatial patterns of damage. The Predictive Risk Analytics Sub-Agent applies machine learning classification and regression methods (e.g., Random Forest, XGBoost) as well as deep learning hazard models to historical hazard records and geospatial datasets such as elevation, land use, and demographics, producing fine-grained risk maps that combine physical and socio-economic dimensions. The Situational Awareness Sub-Agent fuses natural language processing of social media streams with computer vision applied to UAV or satellite imagery, integrating multiple modalities of real-time data to construct a comprehensive operational picture for responders. The Resource Allocation and Logistics Sub-Agent uses optimization techniques (e.g., mixed-integer programming, heuristics) and reinforcement learning for dynamic routing, leveraging forecast data and logistics inventories to plan supply distribution and adapt to disrupted transportation networks. The Evacuation Planning Sub-Agent incorporates reinforcement learning and graph-based flow models to simulate and optimize evacuation routes and shelter allocations, dynamically adjusting recommendations as hazard conditions evolve or road networks are disrupted. Finally, the Misinformation Control Sub-Agent employs NLP models such as BERT and RoBERTa for rumor detection, complemented by graph-based diffusion analysis to identify and contain the spread of false or misleading information across social media, thereby supporting accurate and trusted communication with affected populations. Together, these specialized sub-agents provide modular yet interoperable intelligence that equips the Disaster Copilot system to support decision-making across all critical phases of disaster management.

### 3.2.2 AI-Model sub-planner for step-level selection

To fully exploit these specialized tools, each sub-agent includes an AI-model sub-planner that chooses the appropriate model for a given input context. Within each sub-agent, an AI-model sub-planner operates to coordinate the execution of models at the step level, ensuring that the most appropriate analytical pathway is selected. While the central task planner determines which sub-agent should handle a broader sub-task (e.g., damage assessment or logistics planning), the sub-planner manages how that sub-agent executes the task internally. It achieves this by interpreting the incoming request, evaluating the availability and suitability of candidate models, and sequencing the required operations, such as preprocessing data, invoking one or more models, and post-processing outputs. The sub-planner leverages lightweight decision policies, such as rule-based selectors, confidence thresholding, or reinforcement-learned policies, to decide among multiple candidate models. For example, in a damage assessment sub-agent, the sub-planner may route satellite imagery to a convolutional neural network if high-resolution images are available, but default to change-detection models when only coarse or outdated imagery exists. By enabling adaptive step-level control, the AI-model sub-planner ensures efficient use of resources, maintains resilience under varying data conditions, and delivers outputs that are contextually optimized for disaster response needs.

### 3.2.3 Collaboration across sub-agents

Although each sub-agent is specialized for a distinct phase of disaster management, effective response requires continuous interaction across sub-agents to generate integrated situational intelligence. Outputs from one sub-agent may frequently serve as inputs for another. For example, flood extent maps produced by the Early Warning Sub-Agent are passed to the Predictive Risk Analytics Sub-Agent for exposure estimation, whose results then inform the Resource Allocation and Logistics Sub-Agent in prioritizing supply distribution. These interactions are enabled through standardized input–output schemas and the coordination of the central task planner, which sequences and validates data exchanges. The distributed memory bank also supports indirect interaction by storing validated artifacts—such as geospatial layers, risk scores, and optimization outputs—that are accessible to all sub-agents as shared context. This design ensures interoperability, reduces duplication of effort, and allows sub-agents to build upon one another's outputs. By facilitating structured cross-agent collaboration, the system transforms isolated AI capabilities into a federated intelligence network capable of delivering timely, multi-faceted decision support in dynamic disaster environments.

### 3.2.4 Modularity and extensibility

Finally, this design is inherently modular. As new AI models emerge, they can be integrated into the relevant sub-agent with minimal disruption. This approach fosters scalability and keeps the broader multi-agent platform future-proof, as each domain sub-agent can evolve independently without compromising the overall federated structure.

### 3.3 External information retrieval augmented generation sub-agent

Beyond the specialized AI models, the system offers an external information retrieval augmented generation (RAG) sub-agent that serves as a knowledge gateway to authoritative disaster management resources. This sub-agent anchors generative outputs to verifiable evidence, reducing hallucination risks while enabling access to continuously updated guidance (A. Singh, Ehtesham, Kumar, & Khoei, 2025). It draws from a multi-source disaster knowledge base that integrates flattened text corpora, knowledge graphs, inverted indices, and structured databases sourced from technical reports, research articles, government directives, and operational guidelines. Such hybrid data representations allow the RAG pipeline to support complementary retrieval methods, including dense semantic search, sparse lexical matching, and graph-based reasoning. The RAG workflow begins with a query processor that interprets user inputs, disambiguates references across multi-turn conversations, and accommodates multi-modal signals (e.g., textual prompts paired with maps or images). Candidate documents are then scored and organized through a retriever–re-ranker pipeline, combining dual-encoder embeddings with cross-attention re-ranking for domain-specific precision. Retrieved evidence is fused with the user query in the generator module, which synthesizes a contextually tailored, citation-backed response. To ensure domain alignment, the sub-agent is trained via supervised fine-tuning (SFT) on disaster-relevant tasks and enhanced through contrastive learning, improving discrimination between closely related yet operationally distinct sources. By anchoring AI-generated outputs to external knowledge repositories, the RAG sub-agent provides traceable, evidence-grounded decision support for tasks such as clarifying evolving legal directives, summarizing emerging research, or validating operational best practices. Its modular design also ensures adaptability: as new data sources become available, they can be ingested into the knowledge base and indexed for retrieval without retraining the core model. This makes the RAG sub-agent not only a safeguard against misinformation but also a continuously evolving reservoir of institutional memory for disaster management.

## 3.4 Quantization and resource-limited deployment

To ensure the system can scale from powerful data centers to edge devices in remote or constrained environments, all key modules undergo quantization and performance optimization. For instance, weights and activations in neural networks might be converted to INT8 or INT4 to drastically reduce the memory footprint and inference latency, with minimal accuracy loss (Hwang, Jang, & Kim, 2025). Further, TensorFlow Lite or similar lightweight frameworks support deployment on edge devices such as mobile phones, tablets, or low-power field laptops (Rashidi, 2022). These enhancements empower disaster responders operating in bandwidth-limited areas or under infrastructure disruptions to still run advanced AI pipelines.

## 3.5 On-device orchestrator

While cloud-based infrastructure may handle large-scale data flows, real-world disaster conditions demand on-device capabilities for immediate action in case of network failures. To fill this role, the proposed architecture includes a Small Language Model (SLM) that acts as an on-device orchestrator. This SLM interprets full-modality user inputs locally, such as camera images of damaged roads or voice messages requesting rescue operations, and issues the appropriate calls to sub-agents or stored models. Because it is quantized and streamlined compared to the central MLLM, the SLM can operate within tight resource constraints (Xu et al., 2024). Importantly, it also integrates with everyday phone functionalities (calls, email, social media apps) to relay urgent alerts, send real-time updates, or even capture fresh data for analysis. This ensures that even if cloud services are unavailable, local teams can still benefit from semi-autonomous planning and accurate AI insights wherever they are needed.

## 3.6 Memory bank with current, short-term, and long-term storage

Finally, the memory bank component handles the longitudinal storage of conversation histories, user-uploaded data, and critical domain knowledge, which is essential to both high-quality interactions and organizational knowledge retention (Zhong, Guo, Gao, Ye, & Wang, 2024). It operates on three tiers: a current memory that feeds directly into the LLM for active conversation context, a short-term memory holding summaries of recent tasks and data, and a long-term memory retaining detailed archives of past interactions and user uploads. This distributed architecture not only circumvents the context-window limits of large models, but also intelligently retrieves relevant past information to maintain continuity across extended or multi-day emergency operations. For example, once the system offloads older chat logs or large PDF data to short-term or long-term memory, it can still fetch them via vectorized search methods if needed for a new user question. By systematically managing these memory stores, the system delivers a cohesive user experience, supports advanced multi-turn conversations, and ensures robust domain knowledge retention by preserving context across staff transitions and enabling continuous learning as new disasters unfold.

## 4. Realistic scenario: multi-agent AI system in action

To illustrate the proposed system's real-world value, we envision its deployment during a major hurricane advancing on a coastal region. The scenario unfolds across the disaster lifecycle, demonstrating how the "Disaster Copilot" integrates specialized sub-agents, facilitates human-AI collaboration, and enhances decision-making from initial warnings to long-term recovery.

### 4.1 Hurricane pre-landfall phase

Envision a category-four hurricane advancing on Harris County, Texas in the U.S., a coastal region which is vulnerable to natural hazards, officials at the multi-agency Emergency Operations Center (EOC) begin interacting with the Disaster Copilot through dedicated dashboards and mobile interfaces. In the phase of early warning and risk analysis, a meteorologist queries the system for the latest forecast. The hazard early warning sub-agent is activated, combining satellite weather data with advanced LSTM rainfall models to predict that southwestern districts face severe river overflows. The system automatically passes these projections to the risk analytics sub-agent, which overlays the flood forecast onto historical data and infrastructure maps. It generates and displays a series of risk maps, highlighting vulnerable levees and neighborhoods, allowing public works engineers to proactively place sandbags and reinforcements where they are most needed. An emergency manager asks the system, "Based on the current track, what is our optimal evacuation plan?" The top-level task planner engages the disaster response and resource allocation sub-agent, which uses real-time traffic data and the flood predictions to model various scenarios and recommend dynamic evacuation routes that avoid likely road closures. Simultaneously, the planner tasks the sub-agent to pre-position critical supplies. The sub-agent uses optimization algorithms to determine the most efficient placement of food, water, and medical kits in shelters just outside the highest-risk zones, dispatching delivery notifications to logistics teams via their mobile devices.

**4.2 Hurricane post-landfall phase**

As the storm makes landfall, the emergency management center is inundated with diverse and chaotic data streams. The Disaster Copilot distinguishes itself by synthesizing this complexity into clear, actionable intelligence for timely decision-making. The situational awareness sub-agent continuously scans social media posts and emergency calls, using NLP classifiers to isolate and map urgent rescue requests. These verified alerts appear on the EOC's central dashboard. An emergency responder uploads pre-storm and live drone imagery of a critical bridge querying assessing damages of this structure, and the central task planner routes the query to the impact assessment and recovery sub-agent. It uses its computer vision models to compare the images, detect structural compromise, and generate a detailed report within minutes, flagging the bridge as impassable. This information is instantly shared with the evacuation and resource allocation sub-agents, which automatically recalculate safe routes for rescue teams and supply convoys. A rescue team in an area with no cellular service uses the on-device orchestrator on their tablet. A member voices a command: "I'm trapped by the flood, please save me". The on-device Small Language Model (SLM) uses the device's GPS to get the location and automatically places an emergency call, relaying the coordinates and situation. The team can also take photos of blocked roads, which are analyzed offline by quantized damage assessment models stored on the device to aid immediate, localized decisions.

Through this scenario, the multi-agent system is shown not as a single tool, but as a cohesive operational partner. It breaks down information silos, automates complex data analysis, and provides a continuous, adaptive framework that empowers human experts to make faster, more informed decisions across the entire disaster lifecycle.

### 4.3 Typical workflows

**(1) Workflow using only a domain-professional sub-agent**

Suppose a user inputs the following query: "The above uploaded images are satellite images before and after a flood in the same place. Please give me a detailed report of the damage levels of all buildings in the image." The system first routes the query and the input images to the central task planner. The planner generates a task plan that specifies the use of a domain-professional sub-agent to accomplish this request. Within the sub-agent, the AI-model sub-planner further decomposes the task into steps. It determines that the disaster assessment model should first be applied to the satellite images in order to estimate the damage levels of individual buildings. Once these results are obtained, the outputs are passed to a Geo-Chat module, which generates a detailed explanation and narrative report describing the overall damage situation.

**(2) Workflow using only the external information retrieval–augmented generation (RAG) sub-agent**

Consider a user query such as: "What is the flood inundation depth in socially vulnerable areas in Houston during Hurricane Harvey?" The task planner assigns this query to the RAG sub-agent, which specializes in retrieving and synthesizing external knowledge. Inside the sub-agent, the query processor manages the reasoning process by combining chain-of-thought planning with RAG steps. First, it searches the domain database for information on socially vulnerable areas in Houston (S1). The results are returned to the LLM, which reasons about the next step. Second, it retrieves data on flood inundation depths in Houston during Hurricane Harvey (S2), again returning results to the LLM for contextual reasoning. Finally, it combines these findings to provide inundation depth estimates specifically for the socially vulnerable areas identified in step S1 (S3). The result is a structured, evidence-grounded answer tailored to the user's query.

**(3) Workflow involving interactions among multiple sub-agents**

Now consider a more complex request: "I want to predict the flood inundation depth of the road network in Houston assuming there will be a hurricane similar to Harvey" (Fig 3). The central task planner decomposes this query into multiple steps requiring collaboration among sub-agents. First, it directs the RAG sub-agent to retrieve historical data on flood inundation depths in Houston during Hurricane Harvey. Once these data are collected, they are provided to a domain-professional sub-agent responsible for hydrological prediction. Within that sub-agent, the AI-model sub-planner selects and applies a flood inundation prediction model to generate estimates of water depths under a Harvey-like scenario. These predicted depths are then mapped onto the road network, producing actionable intelligence on potential disruptions to transportation infrastructure.

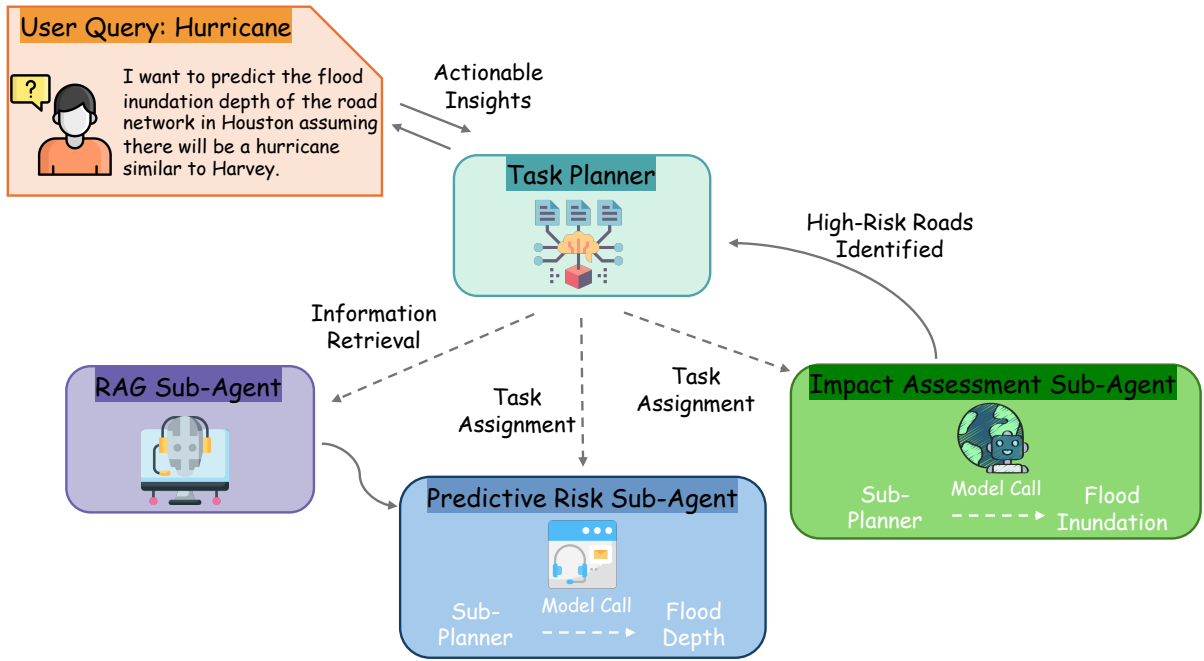

**Fig 3. Multi-agent workflow for flood impact prediction in the Disaster Copilot system.** The figure illustrates how the task planner, powered by a multi-modality large language model, decomposes a complex user query into coordinated subtasks managed by specialized sub-agents. In this example, the user asks to predict flood inundation depths for Houston's road network under a hurricane scenario similar to Harvey. The task planner assigns the RAG sub-agent to retrieve relevant historical flood data, which are then provided to the predictive risk sub-agent for hydrological modeling and estimation of flood depths. These predictions are passed to the impact assessment sub-agent, which evaluates the effects of inundation on transportation networks to identify high-risk roads. The task planner integrates the outputs from each sub-agent to deliver actionable insights back to the user, demonstrating a coordinated, evidence-grounded reasoning process across the Disaster Copilot's multi-agent ecosystem.

## 5. Three-phased roadmap for realizing the multi-agent AI system

Implementing the Disaster Copilot system for disaster management calls for a synchronized evolution across three key dimensions: organizational capacity (O), research & development (R), and human-AI collaboration (H). Fig 4 shows is a three-phased roadmap detailing how these elements can grow together, culminating in a system that seamlessly integrates powerful AI tools with the critical judgment and creativity of human teams.

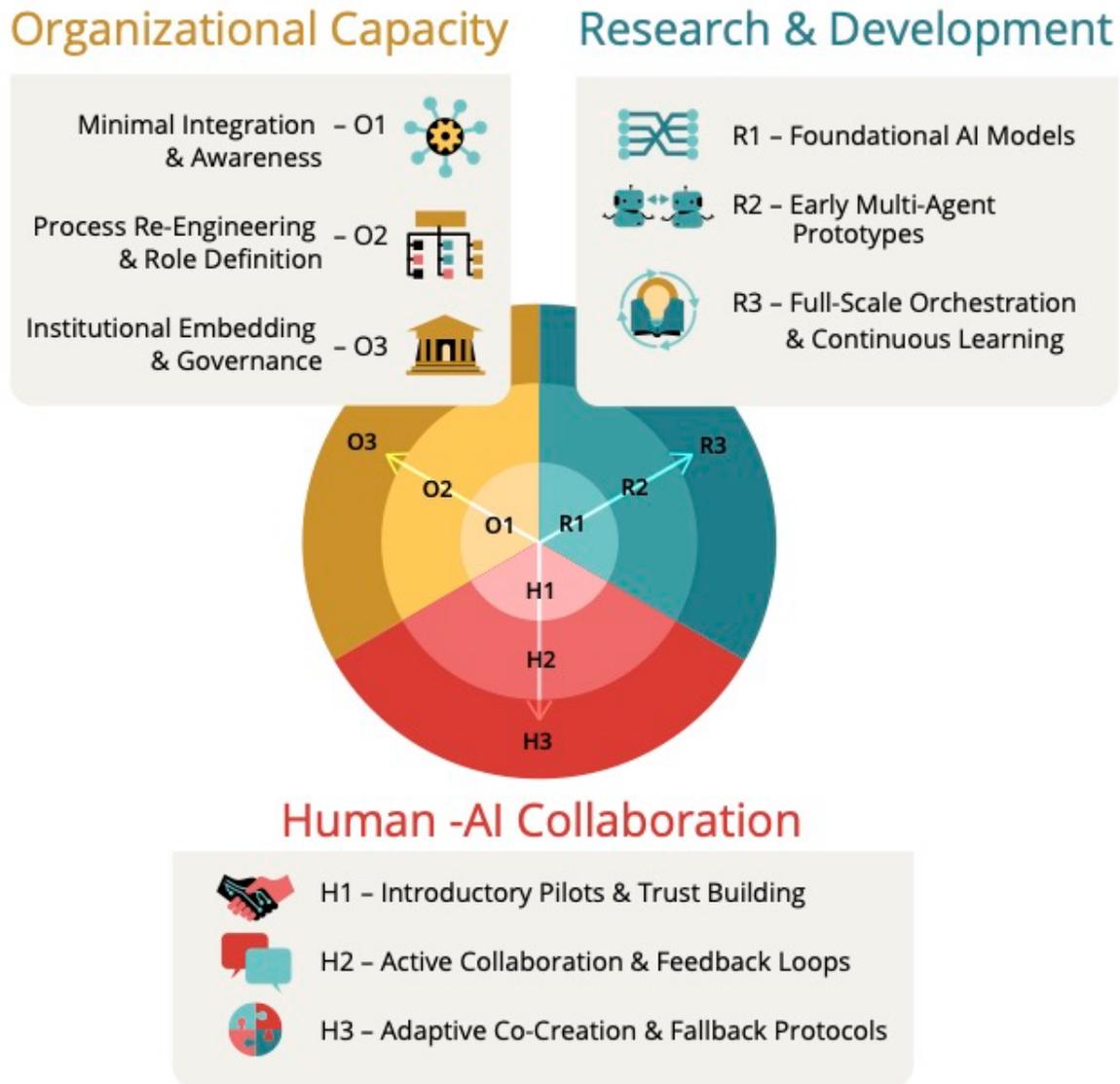

**Fig 4. Three-phased road map for implementing the Disaster Copilot system.** With the synchronized evolution across research & development (R), organizational capacity (O), and human-AI collaboration (H) dimensions: Phase 1 focuses on foundational AI models (R1), basic integration and governance (O1), and initial trust-building exercises (H1); Phase 2 advances to early multi-agent prototypes with orchestration (R2), process re-engineering for AI-embedded workflows (O2), and active coordination with feedback loops (H2); Phase 3 achieves full-scale orchestration with continuous learning (R3), institutional embedding through updated guidelines and training (O3), and adaptive co-creation emphasizing human oversight in dynamic scenarios (H3), ultimately fostering a mature ecosystem for enhanced disaster resilience.

## 5.1 Phase 1: foundational AI models, basic organizational integration, and early human-AI trust

The first phase lays the groundwork for both technological maturity and organizational acceptance of AI. On the research and development front (R1), efforts initially concentrate on creating and validating single-purpose AI models through small-scale pilot implementations to establish baseline accuracy and reliability. However, at this early stage, the models' relative isolation often heightens concerns regarding their opacity and limited interpretability (Visave, 2025). Emergency management teams, therefore, emphasize explainability mechanisms, such as feature attribution and input–output visualization, to mitigate these concerns and help users understand how AI-generated outputs are derived. This focus on transparency is essential for cultivating trust and ensuring that model predictions can be meaningfully interpreted within operational decision-making contexts. In parallel, organizational capacity (O1) initiate basic AI integration measures. This might include holding introductory workshops for department heads, IT personnel, and field officers, emphasizing potential AI benefits and outlining modest data-sharing protocols. An AI integration task force typically forms, bridging the gap between technical developers and operational managers. These early coordination efforts establish the institutional and procedural foundation for a more robust data and workflow pipeline. The task force might also propose minimal governance guidelines, such as clarifying which datasets are permissible to use, ensuring local privacy laws are followed, and specifying who has the authority to adopt or reject AI suggestions in real situations. Meanwhile, the human–AI collaboration dimension (H1) centers on establishing foundational trust between practitioners and emerging AI systems. During this phase, responders and planners engage with prototype tools in which AI outputs operate in an observer mode (Owen, Walker, & Brooks, 2025). Such controlled exposure allows personnel to evaluate the concordance between AI-generated forecasts and conventional meteorological assessments, thereby identifying specific contexts where the AI offers added analytical value. Equally important, these exercises train staff to critically assess and, when necessary, override AI recommendations based on local expertise, reinforcing human oversight as the ultimate locus of decision authority. By the end of phase 1, agencies have a few credible AI sub-agents, a minimal structure for cross-department collaboration, and an initial layer of user familiarity. Although sub-agents are not deeply embedded in daily operations, they have proven their viability, setting the foundation for a broader integration.

## 5.2 Phase 2: early multi-agent prototypes, process re-engineering, and active human-AI coordination

In phase 2, the focus of research & development shifts from single AI components to integrated multi-agent prototypes (R2). At this stage, developers integrate two or three specialized sub-agents under an initial orchestration framework designed to manage data exchange and task sequencing among components. The orchestrator coordinates inter-agent communication, ensuring that outputs from one analytical module seamlessly inform the inputs of another. Preliminary explainable AI interfaces accompany these prototypes, enabling users to visualize the rationale behind orchestration decisions and trace how specific analytical pathways are selected. When benchmarked in realistic operational contexts, these systems serve as testbeds for evaluating interoperability, adaptive behavior under dynamic data conditions, and the transparency of AI-driven coordination. Simultaneously, the agencies realize they must re-engineer existing processes (O2). Rather than treating AI as a standalone advisory, the agencies revise their current operating procedures to integrate sub-agent alerts at specific decision points. For example, an emergency operations center might update its protocol so that evacuation decisions are triggered not just by manual weather station checks, but also by threshold breaches in the AI's predictive model. The Human-AI collaboration dimension (H2) deepens as staff adopt active coordination with these prototypes. Ongoing scenario exercises permit operators to see integrated sub-agent outputs in a single dashboard. Moreover, user feedback loops become standard. If a sub-agent keeps flagging false positives, staff mark it as erroneous, prompting developers to recalibrate model thresholds. Through this iterative cycle, the multi-agent system steadily improves in alignment with field realities. By the end of phase 2, the architecture supports routine usage in recurring disaster events. The operating procedures of agencies can also be shaped around the AI's abilities, systematically weaving human oversight and AI data processing together.

## 5.3 Phase 3: full-scale orchestration, institutional embedding, and adaptive co-creation

The final phase aims for comprehensive multi-agent orchestration (R3), encompassing numerous sub-agents linked under a sophisticated orchestrator. At this time, the system handles multiple hazards in parallel. Developers also finalize continuous learning features that refine sub-agent models after each major event, feeding new data points and user annotations back into the training pipeline. Within organizational capacity dimension (O3), the multi-agent system becomes a structural pillar of emergency management. Emergency agencies adopt it widely by modifying guidelines and procedures around sub-agent usage. Moreover, staff training programs formally incorporate multi-agent system coordination as a core competency for new recruits, ensuring that operational proficiency and institutional knowledge are continuously transferred and preserved despite personnel turnover. As for human-AI collaboration dimension (H3), teams achieve adaptive co-creation with the multi-agent system. Staff trust the AI to handle complex analytics, while they focus on strategic, ethical, or contextual aspects. If contradictory data arises, staff can rapidly challenge the orchestrator's conclusions, leading to a short negotiation loop among sub-agents and operators. Through repeated real events learned, the Disaster Copilot system fosters a cycle of continuous improvement, eventually achieving a mature human-AI dynamic that is both speedy and responsive to ground truths. By aligning all three dimensions within the three-phase implementation map, the multi-agent AI system integrates seamlessly into disaster management, driving data-informed insights and enhancing the creativity of human responders. This synergy advances the vision of disaster response that is faster, more precise, and more adaptable, bringing it significantly closer to reality.

## 6. Discussion

The vision for the Disaster Copilot system represents a paradigm shift for disaster management, which moves from a collection of isolated tools toward an integrated, intelligent ecosystem. This section discusses how the proposed architecture addresses systemic challenges in the field, its theoretical and practical implications, and the limitations and future work required to bring it to fruition.

### 6.1 Multi-agent AI for addressing systemic challenges in disaster management

Many of the persistent challenges in disaster management, such as fragmentation, staff turnover, and limited time to consolidate lessons between successive disaster, stem from deeper structural issues. The proposed Disaster Copilot system goes beyond merely improving individual tasks to directly mitigate these systemic weaknesses. By integrating specialized sub-agents under a central planner and shared memory architecture, the system effectively preserves institutional knowledge, fosters real-time collaboration across agencies, and adapts continuously to new data, even under extreme time pressures.

A key challenge in disaster management is the fragmentation of data and decision-making. Each disaster management agency maintains its own tools and information, leading to siloed operations (Jayawardene, Huggins, Prasanna, & Fakhruddin, 2021). The multi-agent AI system addresses this by federating diverse functionalities. One sub-agent can specialize in real-time satellite image analysis for damage assessment, another in route optimization for evacuation, and yet another in logistics for resource distribution. All these insights flow into a unified top-level planner, ensuring that critical updates are instantly available to every relevant sub-agent and decision-maker. Moreover, the system's knowledge base and long-term memory reduce the dependence on manual information sharing, enabling data from one domain to benefit others without bureaucratic hurdles or communication delays.

Another structural difficulty arises from staff turnover and the consequent erosion of institutional memory. Local agencies dealing with floods, hurricanes, or wildfires often lose key personnel and the hard-earned lessons of previous disasters with them. Our proposed Disaster Copilot system stores operational decisions, annotated outcomes, and relevant after-action analyses in a dedicated memory bank. Over time, this establishes a virtual repository of best practices and hazard profiles. As new staff come on board, they can quickly retrieve relevant decisions from similar past events. By converting transient human expertise into a structured and enduring knowledge asset, the system enables agencies to preserve and reuse collective institutional intelligence across successive staff transitions.

Finally, the system is well-suited for resource-constrained contexts, where staff manage multiple roles with limited budgets. Because many agencies lack the infrastructure for large on-premise computing or reliable network connectivity, the proposed Disaster Copilot architecture incorporates quantized inference and on-device deployment capabilities. These features enable field teams and smaller municipalities to selectively adopt and execute only the sub-agents most relevant to their operational needs, thereby extending advanced analytical functionality to resource-constrained environments. This modular approach enables an incremental rollout, encouraging gradual scaling that helps agencies assimilate each sub-agent's functionality in a way that meets local needs.

**6.2 Theoretical contributions and implications**

This study offers several contributions to theory across information systems, knowledge management, and human-computer interaction. First, this study moves beyond the view of AI as a passive decision-support tool and conceptualizes a multi-agent system in which an orchestrator coordinates specialized sub-agents under human oversight. This architecture provides a theoretical blueprint for collective intelligence in complex, high-stakes socio-technical systems, where effectiveness depends not only on computational power but also on the structured integration of human judgment and machine reasoning.

Second, the study offers a novel model for overcoming technological fragmentation by introducing a multi-agent architecture. This contributes to information systems research by demonstrating how heterogeneous models and data sources can be federated without loss of specialization. The emphasis on standardized interfaces, composability, and provenance tracking also enriches the literature on system interoperability and accountability, providing an alternative to monolithic, centralized AI platforms.

Third, the design of a distributed memory bank and knowledge-capturing sub-agents constitutes a concrete mechanism for mitigating the erosion of institutional memory in high-turnover environments. This operationalizes concepts from knowledge management theory by showing how tacit and explicit knowledge can be systematically captured, stored, and reapplied across successive disaster events. In doing so, the study extends the theoretical discourse on organizational learning and knowledge continuity under conditions of crisis and uncertainty.

### 6.3 Practical implications

The adoption of the *Disaster Copilot* system carries significant practical implications for disaster management practitioners. First, one of the most immediate impacts of the system is its ability to integrate fragmented data and workflows. By ensuring that critical insights, such as flood forecasts or infrastructure damage reports, are instantly available to all relevant stakeholders (e.g., evacuation planners, logistics coordinators), the system supports the development of a shared operational picture. This capability fosters more coordinated, timely, and evidence-based decision-making across agencies that have traditionally operated in isolation. Second, the system is not designed as an add-on technology but as a capability that requires co-evolution with existing operating procedures. Agencies may need to adapt their workflows to formally incorporate AI-generated outputs at key decision points. This implies a gradual but fundamental transformation of organizational routines, where human expertise and AI recommendations are jointly embedded in decision-making protocols. Third, effective use of the system requires personnel to develop new competencies, such as the ability to interpret system-generated confidence scores, provenance indicators, and structured outputs. In some cases, new organizational roles may emerge, who bridges the gap between technical outputs and strategic command decisions. Over time, this evolution may foster a workforce that is not only technologically literate but also better equipped to critically evaluate and collaborate with AI systems in high-stakes disaster contexts. Accordingly, this work provides a vision for disaster AI from isolated tools to a federated, multi-agent system that operators can trust and deploy. Concretely, we specify a task-planning orchestrator that composes specialized sub-agents through standardized interfaces; a multilingual, multimodal interaction layer that fuses text, vision, audio, and geospatial data; a tiered memory bank that preserves institutional knowledge across events and staff turnover; and edge-ready deployment via quantization and an on-device orchestrator for degraded-connectivity settings. Coupled with retrieval-augmented generation and explicit provenance/audit hooks, the system turns fragmented analytics into traceable, actionable decisions. A three-phase roadmap with measurable KPIs and gates translates the architecture into an adoption path for agencies. This architecture operationalizes federated, auditable, and multilingual multimodal AI for EOCs. It is practical under bandwidth constraints and preserves institutional memory, and also offers a measurable adoption path.

### 6.4 Limitations and future directions

While the proposed multi-agent AI framework presents a promising roadmap for transforming disaster management, several limitations highlight areas for further exploration. First, current sub-agents may become brittle when confronted with conditions beyond their training distribution, making model adaptability through online learning and incremental retraining a critical future direction. Second, the system's outputs, often tied to life-critical decisions, also expose the limitation of explainability. Simple confidence scores are insufficient for complex recommendations, and future research should develop tiered explanation mechanisms that tailor detail to different user roles while incorporating real-time operator feedback. Furthermore, the absence of a universally adopted communication protocol remains a structural barrier; developing and operationalizing standards such as the Model Context Protocol (MCP) will be essential to ensure interoperability, traceability, and resilience in low-connectivity environments. Addressing these challenges will require interdisciplinary work across AI, human–computer interaction, and organizational science, ultimately moving the system from conceptual vision toward a robust, field-deployable capability that can continuously evolve with the demands of an increasingly complex risk landscape.

## 7. Concluding remarks

This paper presents a vision for a multi-agent AI paradigm to transform disaster management from fragmented tools into an integrated, intelligent ecosystem. The proposed architecture is composed of four core components: a central task planner that orchestrates complex user queries into actionable workflows, domain-specific sub-agents that encapsulate specialized AI models for tasks such as hazard mapping, risk analytics, and evacuation planning, a federated multi-modality and multilingual interface that broadens accessibility across diverse data streams and communities, and a distributed memory system that preserves institutional knowledge and ensures continuity across successive crises. Together, these elements enable collective human-machine intelligence that is adaptable, transparent, and resilient under real-world disaster constraints. The convergence of multi-agent systems, large language models, and domain-specific AI holds significant promise for reconfiguring disaster management practices, fostering interdisciplinary collaboration, and ultimately building more adaptive and resilient communities.

Theoretically, this study advances the paradigm of disaster management by conceptualizing AI not merely as a collection of isolated tools, but as an active, agentic ecosystem. The proposed Disaster Copilot architecture provides a novel blueprint for achieving collective human-machine intelligence in high-stakes, socio-technical systems, demonstrating how specialized computational expertise can be dynamically orchestrated under human oversight. By prioritizing a federated, multi-agent approach over monolithic platforms, the framework offers a robust theoretical model for overcoming systemic technological fragmentation and enhancing interoperability without sacrificing domain specialization. Furthermore, the system operationalizes key concepts from knowledge management theory, introducing mechanisms such as the distributed memory bank to systematically capture and retain institutional knowledge, thereby ensuring organizational learning and continuity despite the persistent challenges of high staff turnover and successive crises.

Practically, the Disaster Copilot system promises to significantly enhance the agility and effectiveness of disaster response operations. Its primary impact lies in dismantling operational silos by unifying disparate data streams and specialized AI tools into a cohesive, real-time common operating picture. This integration directly accelerates and sharpens the decision-making cycle. For instance, insights from a flood forecasting sub-agent instantly inform the evacuation planning sub-agent, ensuring agile, evidence-based protective actions and eliminating critical delays. By automating complex analytical workflows like damage assessment and resource allocation, the system empowers practitioners to focus on high-level strategic challenges rather than data reconciliation. Crucially, the architecture's emphasis on quantization and on-device orchestration ensures that advanced analytical capabilities are accessible even in resource-constrained or connectivity-compromised environments, significantly bolstering frontline resilience.

Another significant contribution of this study lies in advancing the application of Digital Twin technologies in disaster management. While Digital Twins provide essential virtual representations of physical environments and infrastructure, they often function primarily as sophisticated visualization platforms rather than truly intelligent systems. Their operational utility during a rapidly evolving crisis is frequently constrained by a reliance on manual interpretation and an inability to autonomously synthesize real-time, multi-modal data streams. The Disaster Copilot framework addresses this critical deficit by providing the essential AI backbone, transforming the Digital Twin from a passive representation into an active, agentic environment capable of continuous reasoning and adaptive response.

By embedding the proposed multi-agent architecture within a Digital Twin ecosystem, the Disaster Copilot enables the virtual environment to dynamically model complex interdependencies and cascading failures. Infrastructure damage identified by one specialized agent can instantly trigger updates to logistics simulations and evacuation modeling managed by others. This integration elevates the Digital Twin beyond mere situational awareness, establishing it as a comprehensive platform for predictive scenario testing and optimized response simulation. Ultimately, the synergy of the Disaster Copilot's adaptive reasoning with the Digital Twin's environmental modeling realizes the vision of a truly integrated disaster management system, essential for fostering proactive resilience in an increasingly complex risk.

## Data availability

There is no data availability for this manuscript.

## Code availability

There is no code availability for this manuscript.

## Acknowledgements

This research did not receive any specific grant from funding agencies in the public, commercial, or not-for-profit sectors.

## Competing interests

The authors declare no competing interests.

Jayawardene, V., Huggins, T. J., Prasanna, R., & Fakhruddin, B. (2021). The role of data and information quality during disaster response decision-making. *Progress in disaster science, 12*, 100202.

Jeddi, A. B., Shafieezadeh, A., & Nateghi, R. (2023). PDP-CNN: A deep learning model for post-hurricane reconnaissance of electricity infrastructure on resource-constrained embedded systems at the edge. *IEEE Transactions on Instrumentation and Measurement, 72*, 1-9.

Lakshmanaswamy, P., Sundaram, A., & Sudanthiran, T. (2024). Prioritizing the right to environment: Enhancing forest fire detection and prevention through satellite data and machine learning algorithms for early warning systems. *Remote Sensing in Earth Systems Sciences, 7*(4), 472-485.

Lewis, R. H., Jiao, J., Seong, K., Farahi, A., Navratil, P., Casebeer, N., & Niyogi, D. (2024). Fire and smoke digital twin–a computational framework for modeling fire incident outcomes. *Computers, Environment and Urban Systems, 110*, 102093.

Li, B., & Mostafavi, A. (2024). Unraveling fundamental properties of power system resilience curves using unsupervised machine learning. *Energy and AI, 16*, 100351.

Li, D. (2022). A data-driven approach to improving evacuation time estimates during wildfires for communities with part-time residents in the wildland-urban interface. *International Journal of Disaster Risk Reduction, 82*, 103363.

Li, L., Jiang, S., Yuan, J., Zhang, L., Xu, X., Wang, J., . . . Xu, J. (2024). From data silos to seamless integration and coordination: a data-asset centric approach to smart hospital facility management. *Engineering, Construction and Architectural Management*.

Liao, Y., Wang, Z., Chen, X., & Lai, C. (2023). Fast simulation and prediction of urban pluvial floods using a deep convolutional neural network model. *Journal of Hydrology, 624*, 129945.

Liu, C., & Mostafavi, A. (2025). Floodgenome: Interpretable machine learning for decoding features shaping property flood risk predisposition in cities. *Environmental Research: Infrastructure and Sustainability, 5*(1), 015018.

Liu, J., Lee, J., & Zhou, R. (2023). Review of big-data and AI application in typhoon-related disaster risk early warning in Typhoon Committee region. *Tropical Cyclone Research and Review, 12*(4), 341-353.

Liu, Y., Lu, X., Yao, Y., Wang, N., Guo, Y., Ji, C., & Xu, J. (2021). Mapping the risk zoning of storm flood disaster based on heterogeneous data and a machine learning algorithm in Xinjiang, China. *Journal of Flood Risk Management, 14*(1), e12671.

Ma, K., He, D., Liu, S., Ji, X., Li, Y., & Jiang, H. (2024). Novel time-lag informed deep learning framework for enhanced streamflow prediction and flood early warning in large-scale catchments. *Journal of Hydrology, 631*, 130841.

Maha Arachchige, S., & Pradhan, B. (2025). AI meets the eye of the storm: Machine learning-driven insights for hurricane damage risk assessment in Florida. *Earth Systems and Environment*, 1-21.

Mathews, M. C., Vickery, J., & Peek, L. (2024). Resource exchange patterns between Voluntary Organizations Active in Disaster (VOADs): A multilevel network assessment to improve disaster response capacity. *International Journal of Disaster Risk Reduction, 108*, 104455.

Moayedi, H., & Khasmakhi, M. A. S. A. (2023). Wildfire susceptibility mapping using two empowered machine learning algorithms. *Stochastic Environmental Research and Risk Assessment, 37*(1), 49-72.